\documentclass{appolb}
\usepackage[dvips]{graphicx}

\begin{document}
\pagestyle{plain}
\newcount\eLiNe\eLiNe=\inputlineno\advance\eLiNe by -1
\title{Analytic solutions of the Riemann problem in relativistic hydrodynamics and their numerical applications}
\author{Patryk Mach
\address{M. Smoluchowski Institute of Physics, Jagiellonian University,\\Reymonta 4, 30-059 Krak\'ow, Poland}}
\maketitle

\begin{abstract}
We present an analytic solution of the Riemann problem for the equations of relativistic hydrodynamics with the ultra-relativistic equation of state and non-zero tangential velocities. A 3-dimensional numerical code solving such equations is described and then tested against the analytic solution.

PACS numbers: 47.75.+f, 47.40.Nm, 47.85.-g
\end{abstract}

\section{Introduction}

Solutions of the hydrodynamical Riemann problem were introduced to the numerical hydrodynamics by Godunov already in 1959 \cite{godunov}. After 50 years their significance is hard to be overestimated. Especially in the numerical relativistic hydrodynamics, most of the so-called high resolution shock capturing schemes is based on the modifications of the original Godunov idea (for a good review on this issue see \cite{living_review}).

In general, by a Riemann problem for a set of hyperbolic partial differential equations we understand an initial value formulation, where the initial data consist of two constant states separated by a discontinuity in the form of a plane surface. An elementary introduction and some theorems on such a general case can be found in \cite{evans}. Since in his original paper Riemann was concerned with equations of motion of the inviscid fluid, it is also customary to refer to the Riemann problem as a problem in hydrodynamics \cite{riemann}.

The special case of the relativistic Riemann problem, where the gas in both initial states is assumed to be at rest with respect to a chosen inertial frame, was first considered by Thompson in \cite{thompson}. The general Riemann problem with arbitrary velocities, but in one dimension only, was solved by Smoller and Temple \cite{smoller_temple} for the ultra-relativistic equation of state and by Mart\'{\i} and M\"{u}ller \cite{marti_mueller} for the perfect gas equation of state. The latter work was generalized to the case of arbitrary initial velocities, also tangent to the initial discontinuity, by Pons, Mart\'{\i} and M\"{u}ller in \cite{pons_marti_mueller}.

At this point the relativistic hydrodynamics differ strongly from its Newtonian counterpart. The solution of the Newtonian Riemann problem is independent of the velocities tangent to the initial discontinuity and the whole problem can be treated in just one spatial dimension. In the relativistic case all components of the velocity couple to other hydrodynamic quantities through Lorentz factors, and even the wave pattern of the solution can depend on the tangent velocity.

A solution analogous to that of Pons, Mart\'{\i} and M\"{u}ller has recently been obtained for ultra-relativistic equation of state by Pi\c{e}tka and the author in \cite{mach_pietka}. Due to the simplicity of the ultra-relativistic equation of state, this solution can be expressed almost entirely in analytical terms, and as such it can be used both to construct and test modern multidimensional numerical codes solving equations of hydrodynamics.

Our original motivation for dealing with solutions of the Riemann problem was to construct a numerical scheme that could be used in simulations of cosmological hydrodynamical perturbations in the radiation-dominated universe. It is however worth pointing out that the issue is not only numerical. Recently, Aloy and Rezzolla used a solution of the Riemann problem with non-vanishing tangential velocities discussed in \cite{pons_marti_mueller} to explain a purely hydrodynamical mechanism boosting astrophysical relativistic jets to large Lorentz factors \cite{aloy_rezzolla}.

The ultra-relativistic equation of state, commonly used in cosmology, has the form $p = k \rho$, where $p$ is the pressure, $\rho$ the energy density, and $k \in (0,1)$ is a proportionality constant ($\sqrt{k}$ can be interpreted as the local sound velocity). On the other hand, most numerical codes are adjusted to equations of state depending explicitly on the baryonic (rest-mass) density $n$ and the specific internal energy $\epsilon$ (satisfying $\rho = n + n\epsilon$). This is, for instance, the case of the standard perfect gas equation of state of the form $p = (\gamma - 1) n \epsilon$, where $\gamma$ is a constant. Although the ultra-relativistic equation of state can be viewed as a limit of the perfect gas equation of state with $n \to 0$, $\epsilon \to \infty$ and $n\epsilon = \mathrm{const}$, a numerical code has to be rewritten in order to evolve a fluid with the ultra-relativistic equation of state.

In this paper we will summarize the analysis given in \cite{mach_pietka}, and then describe the construction of a 3-dimensional relativistic hydrodynamical code with the ultra-relativistic equation of state. We will also show the results of tests of this code against the presented exact solution.

\section{Equations of relativistic hydrodynamics}

Equations expressing the conservation of the energy and momentum in the relativistic hydrodynamics can be written as
\begin{equation}
\label{energy_momentum_cons}
\partial_\mu T^{\mu \nu} = 0,
\end{equation}
where
\[ T^{\mu \nu} = (\rho + p)u^\mu u^\nu + p \eta^{\mu \nu} \]
is the energy-momentum tensor of the perfect fluid. In this formula $\rho$ denotes the energy density, $p$ is the pressure, $u^\mu$ are the components of the four-velocity of the fluid, and $\eta^{\mu \nu} = \mathrm{diag}(-1,+1,+1,+1)$ is the metric tensor of the Minkowski space-time\footnote{We will work in Cartesian coordinates in this paper. We will also assume a convention in which Greek indices refer to the space-time dimensions $\mu = 0, 1, 2 , 3$ while Latin ones are reserved for spatial dimensions only $i = 1, 2, 3$.}.

For barotropic equations of state of the form $p = p(\rho)$ the above equations constitute a complete set of relevant equations of hydrodynamics. However, if the equation of state depends explicitly on the rest-mass density $n$ (which is the case for the perfect gas equation of state) we also have to take into account the continuity equation, i.e.,
\begin{equation}
\label{baryonic_density_cons}
\partial_\mu(nu^\mu) = 0.
\end{equation}
Thus, the set of the equations of hydrodynamics is different for those two cases, and this difference becomes especially important if we allow for discontinuous solutions. In the following we will restrict our attention to the ultra-relativistic equation of state of the form $p = k \rho$.

Since we aim at solving the general Riemann problem, it is convenient to express equations (\ref{energy_momentum_cons}) in the form
\begin{equation}
\label{three_plus_one}
\partial_t \mathbf U + \partial_i \mathbf F^i = 0,
\end{equation}
where
\begin{equation}
\label{u_def}
\mathbf U = \left( (\rho + p) W^2 - p, (\rho + p) W^2 v^1, (\rho + p) W^2 v^2, (\rho + p) W^2 v^3 \right)^T,
\end{equation}
and
\begin{eqnarray*}
\mathbf F^i & = & \left( (\rho + p) W^2 v^i, (\rho + p) W^2 v^i v^1 + \delta^{i 1} p, \right. \\
& & \left. (\rho + p) W^2 v^i v^2 + \delta^{i 2}p, (\rho + p) W^2 v^i v^3 + \delta^{i 3} p  \right)^T.
\end{eqnarray*}
Here $W$ denotes the Lorentz factor $W = u^0$, and $v^i$ are the components of the three-velocity $v^i = u^i/W$ (note that $W = 1/\sqrt{1 - v_i v^i}$).

\section{Solutions of the Riemann problem}

We will now search for the solutions of the Riemann problem for equations (\ref{energy_momentum_cons}). Without loss of generality, we can assume that the initial discontinuity is perpendicular to the $x$ axis, and it is located at $x = 0$. Thus, due to the translational symmetry of the initial data in the directions $y$ and $z$, equations (\ref{energy_momentum_cons}) can be reduced to
\begin{equation}
\label{set}
\begin{array}{lclcc}
\partial_t \left( (\rho + p) W^2 - p \right) & + & \partial_x \left( (\rho + p) W^2 v^x \right) & = & 0, \\
\partial_t \left( (\rho + p) W^2 v^x \right) & + & \partial_x \left( (\rho + p) W^2 (v^x)^2 + p \right) & = & 0, \\
\partial_t \left( (\rho + p) W^2 v^y \right) & + & \partial_x \left( (\rho + p) W^2 v^x v^y \right) & = & 0, \\
\partial_t \left( (\rho + p) W^2 v^z \right) & + & \partial_x \left( (\rho + p) W^2 v^x v^z \right) & = & 0.
\end{array}
\end{equation}

The structure of solutions of the Riemann problem for the set of equations (\ref{set}) is similar to that obtained in the Newtonian case. In general, the initial discontinuity can decay into three kinds of elementary waves (not all of them have to be present in the solution) propagating along the $x$ direction and separated by some constant states. One of those waves, the so-called rarefaction wave, is a smooth self-similar solution of (\ref{set}). The other are two discontinuities: a shock wave and a contact discontinuity. The distinction between them is based on the behavior of the pressure $p$ and velocity $v^x$, which can be discontinuous at a shock wave and have to be equal on both sides of a contact discontinuity. In fact, the only quantities that can exhibit a jump at the contact discontinuity in the case of the ultra-relativistic equation of state are $v^y$ and $v^z$.

Denoting the Riemann states, that is the initial data for $x<0$ and $x>0$ by $L$ and $R$ respectively, we can symbolically write the solution of the Riemann problem as $LR \to L \mathcal W_\leftarrow L_\ast \mathcal C R_\ast \mathcal W_\rightarrow R$. Here the subscript arrows refer to the direction from which the particles of the fluid enter the wave $\mathcal{W}$, which, in turn, can be either a rarefaction wave $\mathcal R$ or a shock wave $\mathcal S$ (four different wave patterns corresponding to $\mathcal W_{\leftarrow(\rightarrow)} = \mathcal S_{\leftarrow(\rightarrow)}, \; \mathcal R_{\leftarrow(\rightarrow)}$ are possible). The symbol $\mathcal C$ denotes a possible contact discontinuity and $L_\ast$, $R_\ast$ are some intermediate constant states.

The exact form of the solution can be found by considering the dependence of the energy density $\rho$ on the velocity $v^x$ behind the waves $\mathcal W_\leftarrow$ and $\mathcal W_\rightarrow$. (The exact form of this function depends on the states $L$ and $R$ in front of the waves.) Since at the contact discontinuity only $v^y$ and $v^z$ can be discontinuous,  we must have $\rho_{L_\ast} (v^x_{L_\ast}) = \rho_{R_\ast} (v^x_{R_\ast})$. It turns out that the solution of this equation allows us to establish the solution in the intermediate states $L_\ast$ and $R_\ast$, and the precise form of $\mathcal W_{\leftarrow(\rightarrow)}$. In the forthcoming sections we will derive the required relations $\rho(v^x)$ for both rarefaction and shock waves.

\subsection{Rarefaction wave}

Let us consider a rarefaction wave, that is a smooth solution of (\ref{set}) depending on $x$ and $t$ through $\xi = x/t$. Under such an assumption the equations (\ref{set}) can be reduced to
\begin{equation}
\label{set2}
\begin{array}{lcl}
\xi \frac{d}{d \xi} \left( (\rho + p) W^2 - p \right) & = & \frac{d}{d \xi} \left( (\rho + p) W^2 v^x \right), \\
\xi \frac{d}{d \xi} \left( (\rho + p) W^2 v^x \right) & = & \frac{d}{d \xi} \left( (\rho + p) W^2 (v^x)^2 + p \right), \\
\xi \frac{d}{d \xi} \left( (\rho + p) W^2 v^y \right) & = & \frac{d}{d \xi} \left( (\rho + p) W^2 v^x v^y \right), \\
\xi \frac{d}{d \xi} \left( (\rho + p) W^2 v^z \right) & = & \frac{d}{d \xi} \left( (\rho + p) W^2 v^x v^z \right).
\end{array}
\end{equation}

Non-trivial solutions to the above homogeneous set of ordinary equations exist only if the Wronskian of the whole system vanishes, i.e., when $\xi$ are the eigenvalues of the Jacobian $\partial \mathbf F^x/\partial \mathbf U$. These eigenvalues can be computed yielding
\begin{equation}
\label{eigenvalues}
\xi_0 = v^x, \;\;\; \xi_\pm = \frac{v^x (1 - k) \pm \sqrt{k} \sqrt{\left( 1 - v_i v^i \right) \left( 1 - v_i v^i k - (v^x)^2 (1 - k) \right)}}{1 - v_i v^i k},
\end{equation}
where the eigenvalue $\xi_0$ is twofold degenerate \cite{mach_pietka}. In the special case with $v^y = v^z = 0$, the eigenvalues $\xi_\pm$ can be expressed as
\[ \xi_\pm = \frac{v^x \pm \sqrt{k}}{1 \pm \sqrt{k} v^x}, \]
where $\sqrt{k}$ can clearly be identified with the local speed of sound (in the linearized picture the eigenvalues of $\partial \mathbf F^x/\partial \mathbf U$ can be interpreted as the speeds of propagation of acoustic signals).

It can be shown next that for $\xi \neq v^x$ one have
\[ \rho^\kappa W v^y = \mathrm{const}, \;\;\; \rho^\kappa W v^z = \mathrm{const}, \]
where $\kappa = k/(1 + k)$. An assumption that $\xi = v^x$ leads us to the contact discontinuity, which will be discussed later. Introducing the tangential velocity $v^t = \sqrt{(v^x)^2 + (v^y)^2}$, we can show that $v^t = a W^{-1}\rho^{-\kappa}$, where $a$ denotes a constant.

At this stage equations (\ref{set2}) can be integrated. For $a = 0$ (no tangential velocities) one obtains
\[ \left( \frac{1 + v^x}{1 - v^x} \right)^{\pm \frac{1}{2}} = C_1 \rho^\frac{\kappa}{\sqrt{k}}, \]
while for non-zero tangential velocities we get
\begin{eqnarray*}
\left( \frac{1 + v^x}{1 - v^x} \right)^{\pm 1} & = & C_2 \left( \frac{1 + \sqrt{1 + (1 - k) a^2 \rho^{-2\kappa}}}{1 - \sqrt{1 + (1 - k) a^2 \rho^{-2\kappa}}} \right)^\frac{1}{\sqrt{k}} \\
& & \times \frac{\sqrt{k} - \sqrt{1 + (1 - k) a^2 \rho^{-2\kappa}}}{\sqrt{k} + \sqrt{1 + (1 - k) a^2 \rho^{-2\kappa}}},
\end{eqnarray*}
where $C_1$ and $C_2$ are integration constants, that can be computed knowing the state in front of the wave (for more details concerning this solution see \cite{mach_pietka}).

\subsection{Shock wave}

The solution in the form of a shock wave can be obtained from the appropriate Rankine--Hugoniot conditions. For the set of equations (\ref{energy_momentum_cons}) they can be expressed as
\[ \left[ \left[ T^{\mu\nu} \right] \right] n_\mu = 0, \]
where $n^\mu$ is the unit vector normal to the surface of discontinuity. The symbol $[[ \cdot ]]$ is used to denote the difference between the limits of a given function at both sides of the discontinuity. As we are primarily interested in establishing the state behind the shock wave basing on the known state in front of it, we assume a notation in which the value of a given quantity in front of the shock wave is denoted with a dash, while an unaltered symbol is reserved for the value behind the discontinuity. In this notation a jump of a given quantity $f$ is denoted as $[[f]] = f - \bar f$ (a similar convention was adopted in \cite{anile_russo}).

For the discontinuity surface being a plane perpendicular to the $x$ axis, the components of the normal vector $n^\mu$ can be written as $n^\mu = W_s (V_s,1,0,0)$, where $W_s = 1/\sqrt{1 - V_s^2}$. Here $V_s$ has the interpretation of the shock wave velocity.

The Rankine--Hugoniot conditions can be now written as
\begin{equation}
\label{rankine_hugoniot}
\begin{array}{lcl}
\left[ \left[ \rho W^2 - \kappa \rho \right] \right] V_s & = & \left[ \left[ \rho W^2 v^x \right] \right],  \\
\left[ \left[ \rho W^2 v^x \right] \right] V_s           & = & \left[ \left[ \rho W^2 (v^x)^2 + \kappa \rho \right] \right],  \\
\left[ \left[ \rho W^2 v^y \right] \right] V_s           & = &  \left[ \left[ \rho W^2 v^x v^y \right] \right],  \\
\left[ \left[ \rho W^2 v^z \right] \right] V_s           & = &  \left[ \left[ \rho W^2 v^x v^z \right] \right].
\end{array}
\end{equation}
For $v^x = v^y = 0$ the only physical solution of the above algebraic set of equations is given by
\[ \rho = \bar \rho \left( 1 + \Theta + \sqrt{(1 + \Theta)^2 - 1} \right), \]
where $\Theta = W^2 \bar W^2 (v^x - \bar v^x)^2/(2 \kappa (1 - \kappa))$. The shock wave velocity $V_s$ can be expressed as
\[ V_s = \left[ \left[ \rho W^2 v^x \right] \right] / \left[ \left[ \rho W^2 - \kappa \rho \right] \right]. \]

For non-zero tangential velocities $v^t$ the speed of propagation of the shock wave can be obtained as the root of the following cubic equation
\begin{eqnarray*}
\left( 1 -\bar v^x V_s \right) \left[ (1 - v^x V_s)(1 - \bar v^x V_s) - \frac{1}{k} (v^x - V_s)(\bar v^x - V_s) \right] & & \\
- (\bar v^t)^2 (1 - v^x V_s) (1 - V_s^2) & = & 0.
\end{eqnarray*}
The value of $V_s$ can always be computed with the help of Cardano's formulae, but a Newton--Raphson scheme may be more efficient in numerical applications. The above equation was derived from (\ref{rankine_hugoniot}) under the assumption that $V_s \neq v^x$. The case with $V_s = v^x$, corresponds to a contact discontinuity.

Having obtained a solution for $V_s$ we can express the value of the energy density behind the shock wave as
\begin{equation}
\label{post_shock_density}
\rho = \frac{\bar \rho \bar W^2 (\bar v^x - V_s)\left[ (1 -(v^x)^2)(1 - \bar v_x V_s)^2 - (\bar v^t)^2 (1 - v^x V_s)^2 \right]}{(v^x - V_s)(1 - v^x V_s)(1 - \bar v^x V_s)}.
\end{equation}
The solution is completed by the expression for the tangential velocity
\[ (v^t)^2 = \frac{(V_s - \bar v^x)^2 \bar \rho^2 \bar W^4 (\bar v^t)^2}{\rho^2 W^4 (V_s - v^x)^2}. \]

\subsection{Contact discontinuity}

Apart form the shock waves described above, equations (\ref{rankine_hugoniot}) admit non-trivial solutions for which $V_s = v^x$, that is, the discontinuity is moving with the flow of the fluid. In this case we have $\bar v^x = v^x (= V_s)$ and $\bar \rho = \rho$. There is, however, no constraint on the tangential velocity, and it can exhibit an arbitrary jump. Such a discontinuity is called a contact one.

W should also note that in this respect the hydrodynamics with the ultra-relativistic equation of state differs qualitatively from that of the perfect gas equation of state. In the latter case a contact discontinuity can also be present due to the jump in the rest mass density and the specific internal energy.

\subsection{Solutions of the Riemann problem}

Whether we are dealing with the rarefaction or a shock wave can be distinguished basing on the ratio of the pressure $\bar p$ in front of the wave to the pressure $p$ behind the wave. For $\bar p > p$ we have a rarefaction wave, while the condition $\bar p < p$ is characteristic for a shock wave \cite{taub}. It follows from the results presented in the preceding sections that this distinction can be translated into a condition concerning the velocity $v^x$.

\begin{figure}[t!]
\begin{center}
\includegraphics[width=100mm]{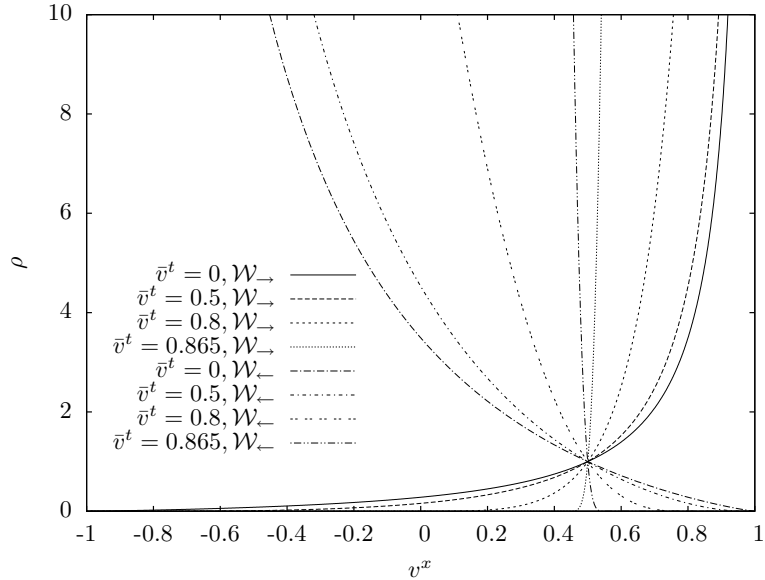}
\vspace{-1em}
\end{center}
\caption{The dependence of the energy density $\rho$ on the velocity $v^x$ behind the wave for the ultra-relativistic equation of state with $k = 1/3$. Different curves refer to values of the tangential velocity $\bar v^t$ in front of the wave equal to 0, 0.5, 0.8, and 0.865. The velocity $\bar v^x$ in front of the wave is equal 0.5, and the density $\bar \rho$ was set to 1. Increasing curves correspond to the right moving waves, while decreasing ones to the left moving waves.}
\label{fig_rho}
\end{figure}

Let $\rho = \mathcal S_{\rightarrow(\leftarrow)}(v^x)$ be the energy density behind the shock wave understood as a function of the post-shock velocity $v^x$ and given by equation (\ref{post_shock_density}) (here again the arrows refer to the direction from which the fluid enters the wave). Let us also introduce a similar function for the rarefaction wave and denote it by $\rho = \mathcal R_{\rightarrow (\leftarrow)} (v^x)$. The expressions for the energy density behind an arbitrary wave $\mathcal W_{\rightarrow(\leftarrow)}$ can be now written as
\[ \rho = \mathcal W_\rightarrow (v^x) =
\left\{ \begin{array}{ll}
\mathcal R_\rightarrow(v^x), & v^x < \bar v^x, \\
\mathcal S_\rightarrow(v^x), & v^x \geq \bar v^x
\end{array} \right. \]
for a right moving wave, and
\[ \rho = \mathcal W_\leftarrow (v^x) = 
\left\{ \begin{array}{ll}
\mathcal S_\leftarrow(v^x), & v^x < \bar v^x, \\
\mathcal R_\leftarrow(v^x), & v^x \geq \bar v^x
\end{array} \right. \]
for a left moving one. The symbol $\bar v^x$ used here denotes the normal velocity in front of a wave. In order to show the influence of the tangential velocities on the solution, we depicted several of such functions on Fig.~\ref{fig_rho}, plotting the graphs for different values of $\bar v^t$ for both right and left-moving waves.

The strategy of finding the solution of the Riemann problem can be now described as follows. We consider a function $\rho = \mathcal W_\leftarrow (v^x)$  for the state $L$ and $\rho = \mathcal W_\rightarrow (v^x)$ for the state $R$, where the states $L$ and $R$ represent the data in front of the waves. The intersection of the graphs of those functions gives the values of $v^x_\ast$ and $\rho_\ast$, common for both intermediate states $L_\ast$ and $R_\ast$, and also identifies the character of both waves (the so-called wave pattern). To complete the solution one only needs to establish the velocities with which the fronts of the waves, and the tail of the rarefaction wave (if present) propagate. The shock wave moves with the velocity $V_s$, which can be easily computed once the value of $v^x_\ast$ has been obtained. The velocity of the head of the rarefaction wave is given by $\xi_\pm$ (plus for $\mathcal R_\rightarrow$, minus for $\mathcal R_\leftarrow$) calculated for the suitable Riemann state. The location of the tail of the rarefaction wave can be computed from the condition that the energy density in the wave should reach the value of $\rho_\ast$.

\begin{figure}[t!]
\begin{center}
\vspace{2em}
\includegraphics[width=100mm]{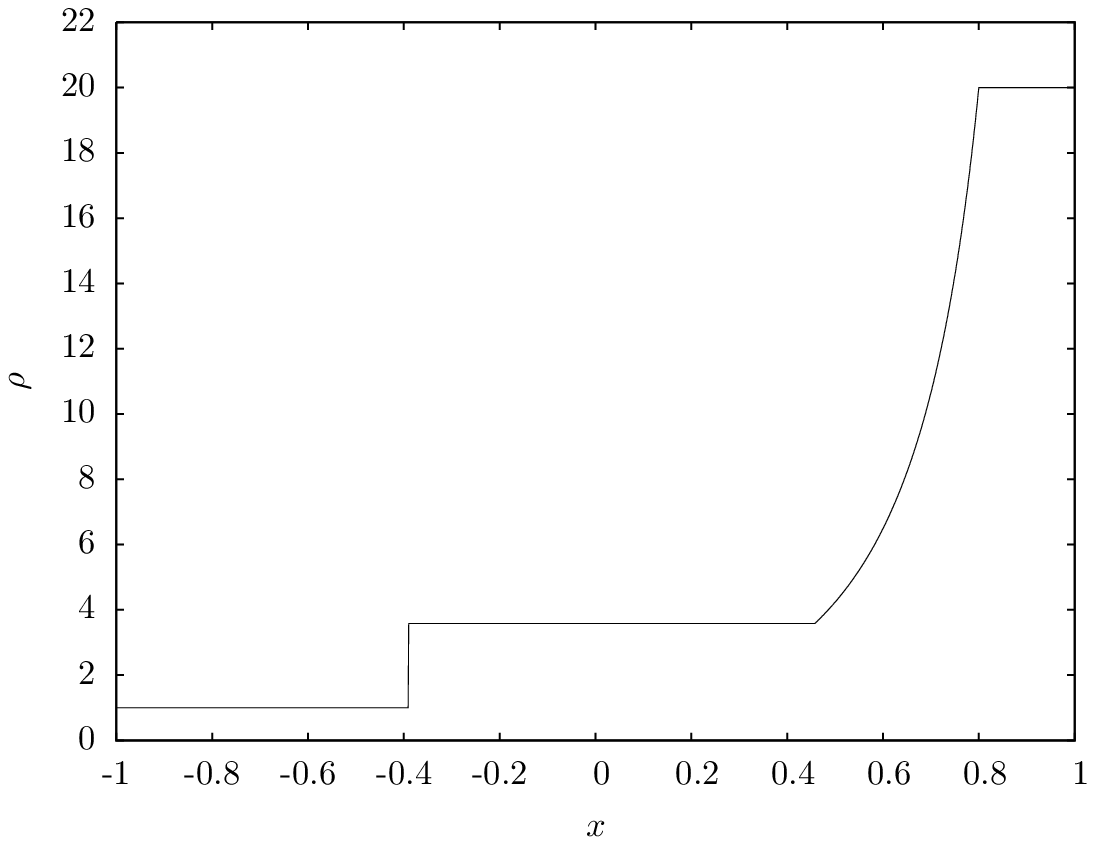}
\vspace{-1em}
\end{center}
\caption{The solution of the Riemann problem for $t = 1$. The left initial state is given by $\rho_L = 1$, $v^x_L = 1/2$, $v^t_L = 1/3$ and the right state by $\rho_R = 20$, $v^x_R = 1/2$, $v^t_R = 1/2$. }
\label{fig_solution2}
\end{figure}

\begin{figure}[t!]
\begin{center}
\vspace{2em}
\includegraphics[width=100mm]{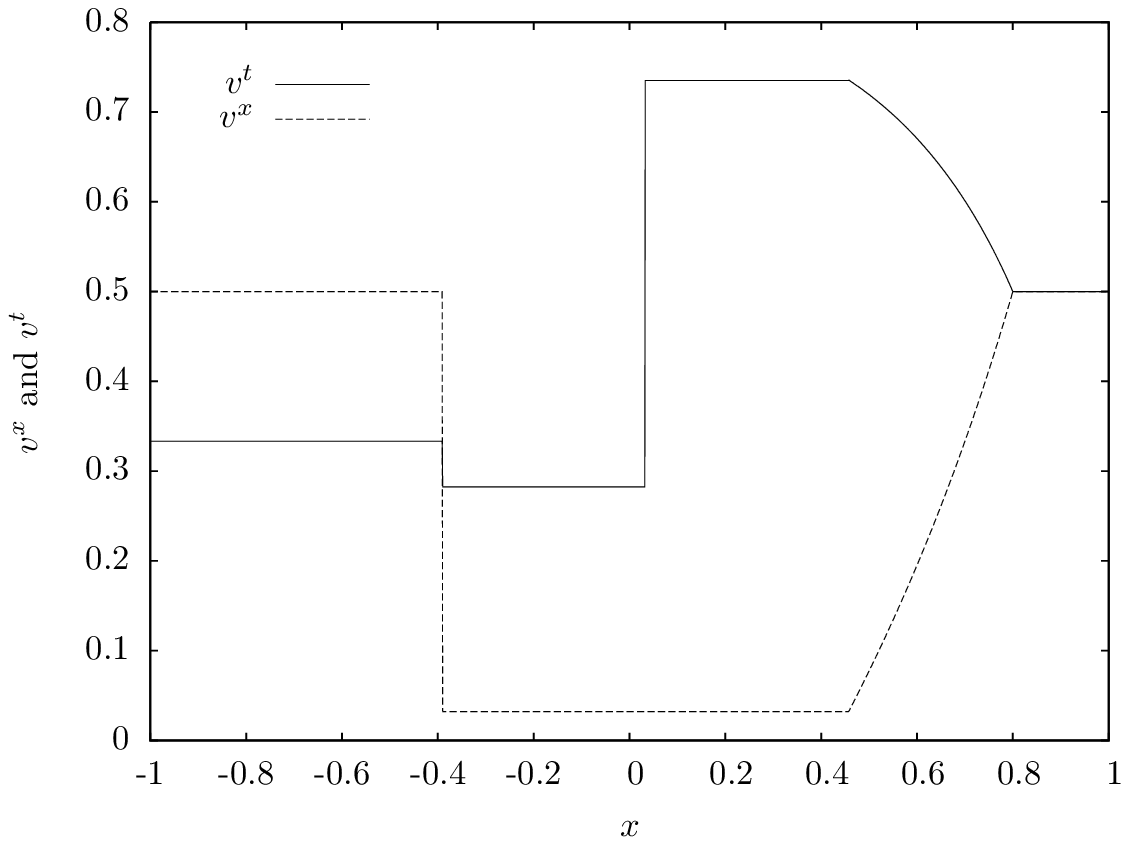}
\vspace{-1em}
\end{center}
\caption{Time snapshot of the solution of the Riemann problem with the same initial data as on Fig.~\ref{fig_solution2}. The solid line corresponds to the velocity $v^t$, while the dotted one depicts $v^x$.}
\label{fig_velocities}
\end{figure}

A sample solution obtained in this way is shown on Figs.~\ref{fig_solution2} and \ref{fig_velocities}. As the initial states for this example we chose $\rho_L = 1$, $v^x_L = 1/2$, $v^t_L = 1/3$ for the left state and $\rho_R = 20$, $v^x_R = 1/2$, $v^t_R = 1/2$ for the right one. The equation of state was assumed to be $p = \rho/3$.

%%%%%%%%%%%%%%%%%%%%%%%%%%%%%%%%%%%%%%%%%%%%%%%%%%%%%%%%%%%%%%%%%%%%%%%%%%%%%%%%
%%%%%%%%%%%%%%%%%%%%%%%%%%%%%%%%%%%%%%%%%%%%%%%%%%%%%%%%%%%%%%%%%%%%%%%%%%%%%%%%
%%%%%%%%%%%%%%%%%%%%%%%%%%%%%%%%%%%%%%%%%%%%%%%%%%%%%%%%%%%%%%%%%%%%%%%%%%%%%%%%

\section{Description of the numerical code}

We will now describe a numerical code that has been implemented in order solve equations (\ref{three_plus_one}) in $(3+1)$ dimensions.

The construction of the code is similar to the one described in \cite{aloy_genesis}. The vector of conserved quantities $\mathbf U$ given by (\ref{u_def}) is discretized on a Cartesian grid of cells and its time derivative is computed according to the following method of lines
\begin{eqnarray*}
\frac{d \mathbf U_{i,j,k}}{dt} & = & - \frac{1}{\Delta x} \left( \hat \mathbf F^x_{i+1/2, j, k} - \hat \mathbf F^x_{i - 1/2, j, k} \right)  - \frac{1}{\Delta y} \left( \hat \mathbf F^y_{i, j + 1/2, k} - \hat \mathbf F^y_{i, j - 1/2, k} \right) \\
& & - \frac{1}{\Delta z} \left( \hat \mathbf F^z_{i, j, k + 1/2} - \hat \mathbf F^z_{i, j, k - 1/2} \right).
\end{eqnarray*}
Here indices $i$, $j$, $k$ number the cells of the grid in the directions $x$, $y$ and $z$, while $\hat \mathbf F^x_{i\pm1/2, j, k}$, $\hat \mathbf F^y_{i, j \pm 1/2, k}$ and $\hat \mathbf F^z_{i, j, k \pm 1/2}$ are numerical fluxes at cells' interfaces. With this discretization the values of $\mathbf U_{i,j,k}$ are evolved with the standard Runge--Kutta method of the second order. The time step for this evolution is limited by the Courant criterion, in which it is convenient to assume the speed of light $c = 1$ as the upper bound for the velocity of any signals.

The numerical fluxes are computed following a method developed by Harten, Lax and van Leer \cite{harten_lax_leer} (HLL method, hereafter) and proposed for the relativistic hydrodynamics by Schneider et al. \cite{schneider_et_al}. A flux between two cells characterized by the states $\mathbf U_L$ and $\mathbf U_R$ is given by
\[ \hat \mathbf F = \frac{c_\mathrm{max} \mathbf F(\mathbf U_L) + c_\mathrm{min}  \mathbf F(\mathbf U_R) - c_\mathrm{max} c_\mathrm{min} \left( \mathbf U_R - \mathbf U_L \right)}{c_\mathrm{max} + c_\mathrm{min}},  \]
where $c_\mathrm{max} = \mathrm{max} \{0, \lambda^{L,R}_0, \lambda^{L,R}_\pm\}$ and $c_\mathrm{min} = - \mathrm{min} \{0, \lambda^{L,R}_0, \lambda^{L,R}_\pm\}$. Here $\lambda_0^{L,R}$ and $\lambda_\pm^{L,R}$ denote the eigenvalues of the Jacobians $\partial \mathbf F^i / \partial \mathbf U$ as given by the formula (\ref{eigenvalues}) and computed for the states $\mathbf U_L$ and $\mathbf U_R$ respectively.

The so-called minmod reconstruction method of Van Leer \cite{leer} was used to improve the spatial accuracy of the scheme. In this approach the states $\mathbf U_{L,R}$ used to compute numerical fluxes are obtained as follows. We first define the ``minmod'' function
\[ \mathrm{minmod}(a,b) = \left\{ \begin{array}{ll} a & \mathrm{if} \;\; |a| < |b|, \;\; ab > 0, \\ b & \mathrm{if} \;\; |a| > |b|, \;\; ab >0, \\ 0 & \mathrm{if} \;\; ab \leq 0 \end{array} \right. \]
and the slope limiters
\[ \mathbf S_i = \mathrm{minmod} \left( \frac{\mathbf U_{i+1} - \mathbf U_i}{x_{i+1} - x_i}, \frac{\mathbf U_i - \mathbf U_{i-1}}{x_i - x_{i-1}} \right). \]
Then the expressions for left and right states used to compute the numerical flux at the interface $i + 1/2$ are given by
\begin{eqnarray*}
\mathbf U^L_{i+1/2} & = & \mathbf U_i     + \mathbf S_i     \left( x_{i+1/2} - x_i \right), \\
\mathbf U^R_{i+1/2} & = & \mathbf U_{i+1} + \mathbf S_{i+1} \left( x_{i+1/2} - x_{i+1} \right).
\end{eqnarray*}
The above formulae actually refer to the reconstruction procedure in $x$ direction. Formulae for other directions are analogous.

The recovery of primitive variables, i.e., obtaining the values of $p$, $\rho$, $v^x$, $v^y$ and $v^z$ from $\mathbf U$ is straightforward, and, unlike in simulations with the perfect gas equation of state, it does not involve a numerical solution to a nonlinear algebraic equation. In the latter case this is usually obtained using an iterative Newton--Raphson scheme. This fact results in a robust performance of the code for the ultra-relativistic equation of state, as compared to the standard case. We omit the exact formulae for the recovery of the primitive quantities, as they follow directly from (\ref{u_def}).

Boundary conditions are implemented by adding to the numerical domain 2 ``ghost'' zones in each direction, following the standard procedure used in numerical hydrodynamics (see e.g. \cite{laney}). The method of ``ghost'' zones is also used to split the grid into parts in the parallel Message Passing Interface (MPI) implementation of the code. In this case the values of the ``ghost'' zone variables are obtained from adjacent parts of the grid.

\section{Sample code tests}

\begin{figure}[t!]
\begin{center}
\vspace{2em}
\includegraphics[width=100mm]{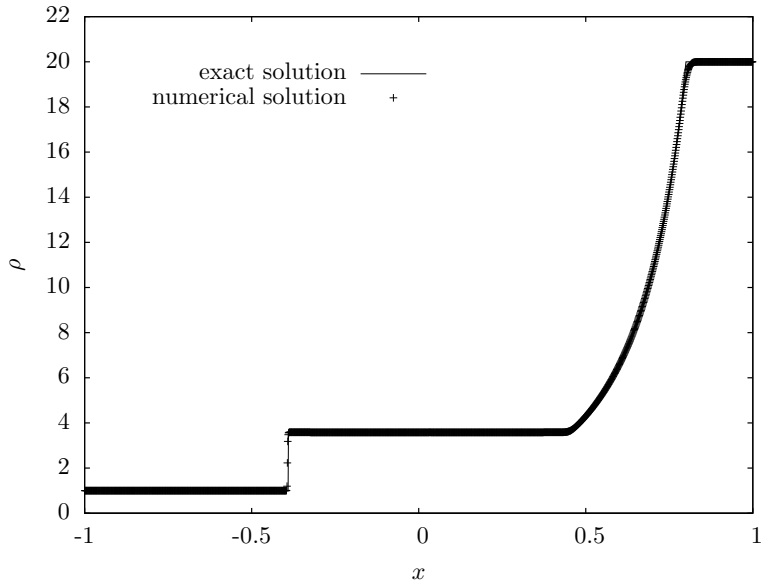}
\vspace{-1em}
\end{center}
\caption{Sample numerical solution for the energy density (points) plotted over the exact solution from Fig.~\ref{fig_solution2} (solid line). The numerical solution was obtained with the spatial resolution of 800 zones per unit length.}
\label{fig_num_density}
\end{figure}

\begin{figure}[t!]
\begin{center}
\vspace{2em}
\includegraphics[width=100mm]{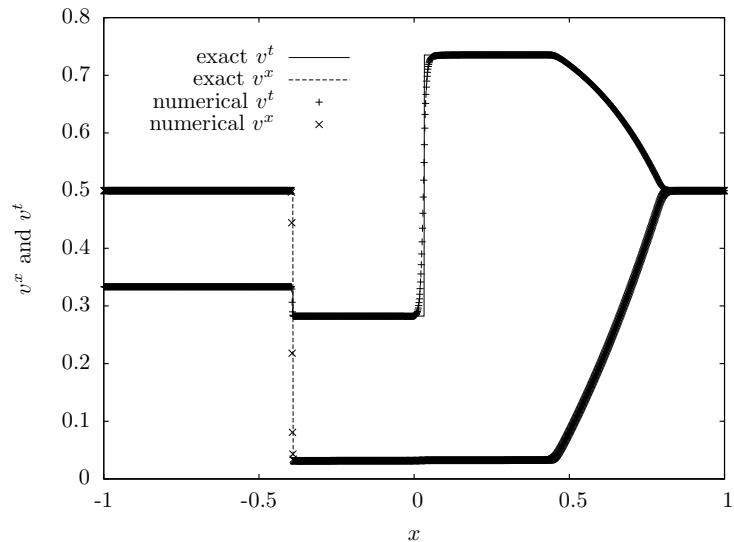}
\vspace{-1em}
\end{center}
\caption{Sample numerical solutions for the velocities $v^x$ and $v^t$ (points) plotted over the exact curves from Fig.~\ref{fig_velocities}. The numerical solutions were obtained with the spatial resolution of 800 zones per unit length.}
\label{fig_num_velocities}
\end{figure}

We used the exact solution presented on Figs.~\ref{fig_solution2} and \ref{fig_velocities} to test the code. To this end we performed a couple of 2-dimensional runs, so that the tangential velocity was also evolved, reaching the time $t = 1$. We used equidistant grids of 100, 200, 400, 800 and 1600 zones per unit length. In each case the Courant factor was set to 0.1. Results of a run with the resolution of 800 zones per unit length are shown on Figs.~\ref{fig_num_density} and \ref{fig_num_velocities}. The data depicted on these graphs come from one $y = \mathrm{const}$ slice of the grid. We see that the resolution of discontinuities is worse for the tangential velocities, most probably due to non-negligible numerical viscosity of the scheme.

Table \ref{tab} gives the estimates of the absolute numerical errors computed with respect to the $L_1$ norm on the interval $x \in [-1,1]$. The exact solution was taken as the reference for the calculation of these errors. The linear convergence in all variables is characteristic for test problem involving strong discontinuities.

\begin{table}[t!]
\begin{center}
\begin{tabular}{cccc}
Zones / unit length & $\rho$ & $v^x$ & $v^y$ \\
\hline
\hline
100  & $3.1 \cdot 10^{-1}$ & $1.6 \cdot 10^{-2}$  & $1.8 \cdot 10^{-2}$ \\
200  & $1.7 \cdot 10^{-1}$ & $9.1 \cdot 10^{-3}$  & $1.0 \cdot 10^{-2}$ \\
400  & $9.2 \cdot 10^{-2}$ & $6.2 \cdot 10^{-3}$  & $6.6 \cdot 10^{-3}$ \\
800  & $4.7 \cdot 10^{-2}$ & $2.8 \cdot 10^{-3}$  & $4.1 \cdot 10^{-3}$ \\
1600 & $2.5 \cdot 10^{-2}$ & $1.7 \cdot 10^{-3}$  & $2.5 \cdot 10^{-3}$
\end{tabular}
\end{center}
\caption{Estimates of absolute errors of sample numerical runs for the Riemann problem illustrated on Figs.~\ref{fig_solution2} and \ref{fig_velocities}. The errors were computed in the $L_1$ norm on the interval $x \in [-1,1]$ with respect to the exact solution.}
\label{tab}
\end{table}

We should also point out that the HLL Riemann solver implemented in the scheme is an approximate one and it does not exploit the analytical solution presented in this chapter. Thus the presented test and the implementation of the code are independent---we are not testing a given solution against itself.

\section{Summary}

We have presented the exact solution of the Riemann problem in the relativistic hydrodynamics with the ultra-relativistic equation of state, in which the fluid is allowed to move with arbitrary velocities, also tangent to the surface of the initial discontinuity. We have also described a 3-dimensional numerical scheme for solving relativistic equations of hydrodynamics with the ultra-relativistic equation of state. Tests of the implemented code against the obtained analytic solution show satisfactory convergence and accuracy allowing to hope for future applications of the code.

We are also convinced that the analytic solution discussed in this article can also be used to construct an exact Riemann solver for the 3-dimensional hydrodynamic codes with the ultra-relativistic equation of state.

We also believe that this solution can find its applications outside strictly numerical hydrodynamics. It was already mentioned in the beginning that a similar solution for the perfect gas equation of state can be used to explain a boosting mechanism present in relativistic jets \cite{aloy_rezzolla}. The required ``boosting'' property is also exhibited by solutions of this paper. This fact can be noticed, for instance, by looking at the solution from Fig.~\ref{fig_velocities}, where the tangential velocity in the rarefaction wave reaches values much greater than any of velocities present in the initial states.

%%%%%%%%%%%%%%%%%%%%%%%%%%%%%%%%%%%%%%%%%%%%%%%%%%%%%%%%%%%%%%%%%%%%%%%%%%%%%%%%
%%%%%%%%%%%%%%%%%%%%%%%%%%%%%%%%%%%%%%%%%%%%%%%%%%%%%%%%%%%%%%%%%%%%%%%%%%%%%%%%
%%%%%%%%%%%%%%%%%%%%%%%%%%%%%%%%%%%%%%%%%%%%%%%%%%%%%%%%%%%%%%%%%%%%%%%%%%%%%%%%

\end{document}